\documentclass[conference]{IEEEtran}

\usepackage{amsmath}
\usepackage{amsfonts}
\usepackage{amssymb}
\usepackage{graphicx}

\ifCLASSINFOpdf
\else
\fi

\newtheorem{lemma}{Lemma}

\newtheorem{theorem}{Theorem}

\newcommand{\FSLD}{{\mathrm{FSLD}}}
\newcommand{\FS}{{\mathrm{FS}}}
\newcommand{\RLU}{{\mathrm{(RLU)}}}

\begin{document}

\title{A Constrained-Dictionary version of LZ78 asymptotically achieves the
Finite-State Compressibility with a distortion measure}

\author{\IEEEauthorblockN{Jacob Ziv}
\IEEEauthorblockA{Department of Electrical Engineering\\
Technion---Israel Institute of Technology\\
Haifa\ 3200003, Israel\\
Email: jz@ee.technion.ac.il}
}


%


\maketitle

\begin{abstract}
\boldmath
The unrestricted-dictionary type LZ78 universal data-compression algorithm
(as well as the LZ77 and LZW  versions)  achieves asymptotically, as the block-length tends to
infinity, the FS compressibility, namely the best compression-ratio that may
be achieved by any Information-lossless(IL) block-to-variable finite-state(FS)
algorithm, for any infinitely-long individual sequence.

One  common practical heuristic  approach  is a constrained-dictionary version
of LZ78, applying the ``Least Recently Utilized'' (LRU) deletion
approach, where only the most recent D entries are kept in the dictionary (denoted by  LZ78(LRU)).

In this note, for the sake of completeness, it is demonstrated again via a
simple proof that the unrestricted LZ78 algorithm asymptotically achieves the
FS-Compressibility. Then, it is demonstrated that the LZ78(LRU)
information-lossless data-compression algorithm also achieves the FS
compressibility, as  the dictionary size D tends to infinity. Although this is
perhaps not surprising, it does nevertheless yield  a theoretical optimality
argument for the popular LZ78(LRU) algorithm (and similarly, for the LZW(LRU) algorithm).

In addition, the finite-state compressibility of an individual sequence under
a constrained allowable distance  measure between the original sequence and
the decompressed  sequence is defined. It is demonstrated that a particular
adaptive  vector-quantizer that sequentially replaces  clusters of L-vectors
onto a single, cluster-representative  L-vector, followed by a constrained
D-entries-dictionary version of LZ78(LRU) as above, is asymptotically
optimal as D tends to infinity and L=$\log$ D.
\end{abstract}


%
\IEEEpeerreviewmaketitle

\section{Introduction and Summary of Results:}
Consider sequences $ x_{1}^{k}$ =$x_{1}, x_{2}, ..., x_{k}; x_{i}\in {\bf A};
i=1, 2, ..., k$
where $|$$ {\bf A}$$|$=A.

Also, let ${\bf x}$=$ x_{1}^{{\infty}}$.

The unconstrained LZ78 universal data compression algorithm has been
introduced in \cite{1}, where it is shown that
when applied to an individual sequence $x_1^n$, as n tends to infinity, it
achieves the FS compressibility.

Let a finite-state encoder be denoted by the triple $({\bf S} , g, f)$ where
$\bf S$ is a finite set of states,
g:$\bf S$x$\bf A$$\rightarrowtail$ $\bf S$, and f:$\bf S$x$\bf
A$$\rightarrowtail$ ${\bf B}^{\star}$, where ${\bf B}^{\star}$ is the set of
all binary sequence.

For each starting  state $s_{1}$, the triple defines a mapping  from ${\bf x}\in {\bf A}^{\infty}$
into $y \in {\bf B}^{\infty}$, where $y_{i}=f(s_{i}, x_{i})$ is a (possibly
empty) binary word, $s_{i+1}=g( s_{i}, x_{i})$ is the next state and where $ i=1, 2, \ldots $.

An information-lossless (IL) finite-state encoder is one for which for each
$n$, the sequence $x_{1}^{n}$ is determined by $y_{1}^{n}$, $s_{1}$ and $s_{n+1}$.

The corresponding compression-ratio for $x_{1}^{n}$ is $ \frac{1}{n\log A}
\sum_{1}^{n} L(y_{i})$, where
$L(y_{i})$ is the length in bits of the (possibly empty) binary word $y_{i}$.

The minimum compression ratio for $x_{1}^{n}$ over all finite-state IL
encoders with at most s states is denoted by $\FS_{s}(x_{1}^{n})$.

Also, let
$\FS_{s}(\bf x) = \limsup_{n \to \infty}  \FS_{s}(x_{1}^{n})$ and let the FS
compressibility of $\bf x$ be defined by,
\begin{equation*}
\FS(\bf x) = \lim_{s \to \infty} \FS_{s}(\bf x).
\end{equation*}

Consider now the parsing of $x_{1}^{n}$ into some c (not necessarily distinct) phrases:
\begin{gather*}
x_{1}^{n}  =  {\bf X}_{1} , {\bf X}_{2} , \ldots , {\bf X}_{j} , \ldots , {\bf X}_{c} ; {\bf X}_{j} =
x_{i(j)}^{i(j+1)-1} ; \\
j=1,2, \, \ldots , \, c .
\end{gather*}
Let ${\bf Z}_{j} ; j=1, 2, \ldots , k; k  \leq  c $ denote the $k$ {\it distinct} substrings among
the $c$ phrases in $x_{1}^{n}$,
where $s_{i,j}$ denotes the start state  and $s_{o,j}$ denotes the end state of the phrase ${\bf Z}_{j}$.

Also, let $L({\bf Z}_{j}|s_{i,j})$ denote the length of the binary
code-word that is generated by the IL FS encoder above,
when fed with ${\bf Z}_{j}$, given the start state $s_{i,j}$.

Let $p({\bf Z}_{j}|s_{i},s_{o})$ denote the empirical probability (i.e.
fraction) of ${\bf Z}_{j}$ among all phrases that are
characterized by a start state $s_{i,j}=s_{i}$ and an end state $s_{o,j}=s_{o}$.

Similarly, let $p({\bf Z}_{j})$ denote the the empirical probability of ${\bf Z}_{j}$
among the $c$ phrases in $x_{1}^{n}$ and let $p(s_{i},s_{o})$ denote the
empirical probability of the pair of states ($s_{i},s_{o}$)
among the  (initial,end) pairs of states of the $c$ phrases.

The corresponding compression-ratio for $x_{1}^{n}$ is
\begin{eqnarray*}
\lefteqn{\frac{1}{n\log A}}\\
&& \hspace*{-.55cm}\sum_{1}^{c}L({\bf Z}_{j}|s_{i})= \frac{c}{n\log A}
\, \sum_{s_{i}=s(1)}^{s} \, \sum_{s_{o}=s(1)}^{s}  p(s_{i},s_{o})\\
&& \hspace*{-.55cm} \sum_{1}^{k}p({\bf Z}_{j}|s_{i}.s_{o})L({\bf Z}_{j}|s_{i}) ,
\end{eqnarray*}
where $ s(t);t=1, 2, \ldots , s   $ are the distinct states that appear at the
start or at the end of any of the $c$ phrases (at most s such states.\
\begin{lemma}

Consider an arbitrary parsing of $x_{1}^{n}$ into c substrings (phrases).
Then,
\begin{gather*}
FS_{s}(x_{1}^{n}) \geq {\frac{c}{n\log A }} \bf \left[ \sum_{1}^{k} ( p({\bf Z}_{j})
\log\left(\frac{1}{p({\bf Z}_{j})}\right) -2\log s) \bf \right]\\
-{\bf{\it O}}\left(\displaystyle\frac {c}{n\log A}\right) .
\end{gather*}
\end{lemma}

{\bf Proof:}

For a given states pair $s_{i}, s_{o}$, an IL FS  encoder outputs a distinct
binary code-word for each of
the $c$ phrases that start with the state $s_{i}$ and end with $s_{o}$.
Observe that all such phrases may be permuted without  changing the
code-length for the whole sequences.
Thus,counting the total number of such code-length preserving permutations

yields  by Stirling formula \cite{6}:
\begin{eqnarray*}
&& \sum_{1}^{c}p({\bf Z}_{j}|s_{i}.s_{o})L({\bf Z}_{j}|s_{i})\\
\geq&& \sum_{1}^{c} p({\bf Z}_{j}|s_{i}, s_{o}) \log\left(\frac{1}{p({\bf
Z}_{j}|s_{i}. s_{o})}\right)-{\bf {\it O}}(\log c)
\end{eqnarray*}

Lemma 1 follows immediately by observing that
$ -\log p({\bf Z}_{j}|s_{i}, s_{o})\geq -\log p({\bf Z}_{j})-\log p(s_{i},
s_{o})$.
Now, in the case of LZ78 \cite{1}, all the $c$ phrases that are generated for
$x_{1}^{n}$, are all distinct (except perhaps of the last phrase). For
example, in the case of the LZ78 algorithm, each new phrase is either an
extension of a previous phrase by one letter, or a single letter that is not
identical to any of the past single-letter phrases. The code length
for each phrase is bounded by
$\log $$C_{n}(\mathrm{LZ78})$+1+$\log A $, where $C_{n}(\mathrm{LZ78})$ is the
number of distinct phrases that are generated by LZ78.

Therefore,
\begin{lemma}
For any individual sequence $\bf x$
\begin{equation*}
\FS({\bf x}) \geq \limsup_{n\to\infty}\frac{1}{n\log A} [(C_{n}(LZ)78)\log(C_{n}(\mathrm{LZ78})] .
\end{equation*}
\end{lemma}

The main result in \cite{1} follows from Lemma~1 and Lemma~2 as follows:

The compression-ratio that is achieved for an individual sequence $x_{1}^{n}$
that is parsed into $C_{n}$ (LZ78) distinct phrases by LZ78 is upper-bounded by
$$\frac{1}{n\log A} \Bigl(C_n(\mathrm{LZ78})\Bigr)( \log C_n (\mathrm{LZ78})+1+\log A).$$

Thus,
\begin {lemma}
The LZ78 universal IL data-compression algorithm asymptotically achieves
$\FS(\bf x)$.
\end{lemma}

Similarly, it follows that Lemma 3 holds for LZW \cite{2}  and LZ77 \cite{3} as well.
In practice, in order to avoid the ever growing size of the dictionary that
contains all the past phrases that are generated by LZ78
(or similarly, by LZW), heuristic  constrained-dictionary versions has been proposed.

Apparently, the preferred heuristics is  the Last-Recently-Used (LRU) method
\cite{4}. In this case, only the most recent
phrases ( no larger than some preset number D) are kept in the dictionary.

This approach is analyzed below, and is shown to asymptotically achieve $\FS(\bf x)$ as well.

Consider a constrain-dictionary LZ78 algorithm, where the dictionary has $D$
entries,
each no longer than $L_{\max} = (\log D)^2$ letters.
Each newly generated phrase is a copy of the longest matched phrase among the
previous $D$ phrases,
extended by the next incoming
letter. If no match is found with any of the phrases in the dictionary, then
the first incoming letter is the next phrase.

The new phrase is then included in the dictionary and the last recently used
phrase is removed from the dictionary, except
for the case where the newly generated phrase is of length $L_{\max} +1$, in
which case the dictionary is not updated.

The code length for each successive phrase is $\log D +1+ \log A$.  Denote
this algorithm by LZ78(LRU).

\begin{theorem}
The compression-ratio that is achieved by LZ78(LRU) when applied to an
individual $\bf x$ converges asymptotically to $\FS(\bf x)$ as $D$ tends to infinity.
\end{theorem}

$\bf Proof$:
Let $c(n)$ denote the number of phrases that are generated by LZ78(LRU) when
applied to $x_1^n$ and let $c(n | L_{\max}+1)$ denote the number of phrases of length $L_{\max} +1$.

By construction, $p({\bf Z}_{j})\leq \frac{1}{D}$ for any phrase $Z_{j}$
among the c(n) phrases that are no longer than $L_{\max}$ since the number of
phrases in between any such phrase and it's most recent previous appearance is
at least $D$ (since it is not included in the dictionary).

Let ${\rho}_{\mathrm{LZ78}\RLU}(x_1^n)= \frac{C(n)}{n\log A} (\log D+1+\log
A)$
denote the compression-ratio that is achieved by LZ78(RLU) when applied to
$x_1^n$.

By Lemma 1,
\begin{eqnarray*}
\lefteqn{\rho_{LZ78(RLU)}(x_1^n) = \frac{C(n)}{n{\log A}}(\log D+1+\log A)}\\
& \leq & FS_{s}(x_{1}^{n}) + {\bf {\it O}}\left(\frac{c(n)}{n{\log A}}(2\log s +1+\log A \right )\\
&& +\frac{1}{n{\log A}}c(n|{L_{\max}}+1)\log D + {\bf {\it O}}\left(\frac { c(n)}{n{\log A}}\right)
\end{eqnarray*}
Therefore,
\begin{eqnarray*}
\lefteqn{\rho_{LZ78(RLU)}(x_1^n) \left(1- \frac{2\log s}{\log D}\right)}\\
& \leq & FS_{s}(x_{1}^{n})+\frac{c(n)|L\leq L_{\max}) }{n\log A}(\log D+1+\log A)\\
&& +{\bf {\it O}}(\frac { c(n)}{n{\log A}}) +\frac{1}{n\log A} c(n |{L=L_{\max}}+1) \log D
\end{eqnarray*}
where $c(n |{L=L_{\max}}+1)$ denotes the number of phrases among the $c(n)$ phrases, of length $L_{\max}+1$.

Observe that $n \geq c(n|L=L_{\max}+1)(L_{\max}$+1) and that $ L_{\max}$=$(\log D)^2$.
Also, by construction, $c(n){\log D}\leq n{\log A}{\rho_{LZ78(RLU)}(x_1^n)}$
and hence, $\frac{c(n)}{n{\log A}}\leq \frac{\rho_{LZ78(RLU)}(x_1^n)}{\log D}$
which proves Theorem~1.

The same result holds for LZW(LRU) as well as for a sliding version of LZ77
where the window is set at $DL_{\max}$
and where the phrase length is constrained to be no larger than $L_{\max}$.

The fact that a sliding window version of LZ77, where the phrase is  not
constrained to be no longer than $L_{\max}$,
yields a compression ratio that is equal to $\FS(\bf x)$  was already
established by P.~Shields \cite{5}.

It should also pointed out that while LZ78 and LZW are {\it not}
finite-state algorithms, LZ78(RLU), LZW(RLU) and the sliding-window version
of the LZ77 algorithm  are all elements of the class for which $ \FS(\bf x)$ is defined.

Now, let $d(x_1^{i};y_1^{i})$ denote some given distance measure between the
vectors $x_1^i$ and $y_1^i$, satisfying:
\begin{align*}
& d(x_1^{i};y_1^{i})+d(x_{(i+1)}^{(i+j)};y_{(i+1)}^{(i+j)})\\
 \geq & d(x_1^{(i+j)};y_1^{(i+j)}) \, ; \; i,j=1,2, \ldots
\end{align*}

Let a finite-state distortion-limited (FSDL) encoder for L vectors  be one
such that
for each starting  state $s_{i}$, and an end state $s_{o}=g(s_{i},
x_{1}^{L})$
it defines a mapping  from ${x_1^L}\in {\bf A}^{L}$ to
$Y(1)\in {\bf B}^{\infty}$,  where $Y(1)=f(s_{1}, x_1^{L})$  is a
(possibly empty) word that, given  the states $s_{i}$ and $s_{o}$ generates
some vector ${z_1^L}\in {\bf A}^{L}$  such that $d(x_1^{L};z_1^{L})\leq Ld_{\max}$.

This typifies cases (e.g. bio-genetics) where any two L-vectors for which
the distance measure between the two vectors is no larger than $Ld_{\max}$ are
declared to be similar.

Consider the case where $x_{1}^{N}$ is a concatenation of $L$ substrings
(phrases), where the length of each phrase is $L$ where $N=cL$ is a multiple of $L$.

The corresponding minimal compression-ratio for $x_{1}^N $ over all FS encoders  with $s$
states that satisfy the $d_{\max}$ condition is denoted by
$ \FSLD_{s}(x_{1}^N;d_{\max}|L) = \frac{1}{{N}\log A}\sum_{m=1}^{c} l({\bf Y}(m))$,
where $l({\bf Y}(m))$ denotes the length of ${\bf Y}(m)$ that is associated with
${\bf X}(m)$ and the minimizing states, where ${\bf X}(m)$
is the m-th L-phrase in the parsed $x_{1}^N$.

The FSLD compressibility of $\bf x$ is defined by:
\begin{eqnarray*}
\lefteqn{\FSLD({\bf x};d_{\max})}\\
&=& \limsup_{s\to\infty}\limsup_{L \to \infty}\\
&& \limsup_{N \to \infty}{\frac{L}{N }} \sum_{m=1}^{c}\FSLD_{s}({\bf X}({m};d_{\max}|L)
\end{eqnarray*}

where $p(m)$ is the empirical probability of ${\bf Y}(m)$.

Thus, similar to Lemma 1 above,
\begin{lemma}
\begin{eqnarray*}
\lefteqn{\FSLD_{s}(x_{1}^N;d_{\max}|L)}\\
& \geq & \frac{L}{N\log A}\sum_{m=1}^{c} p(m){\log\left(\frac{1}{p(m)}\right)} -
\frac{2\log s}{L \log A}-{\bf {\it O}}\left(\frac{1}{L{\log A}}\right)
\end{eqnarray*}
\end{lemma}

where $p(m)$ is the empirical probability of ${\bf Y}(m)$.

Next, we describe an adaptive FS quantizing process for L-vectors, that when combined with
the constrained dictionary  version of LZ78(LRU) that is described above,
asymptotically achieves $\FSLD(\bf x)$.

Strings of length $N$  are sequentially replaced by quantized phrases of length L as  follows:

\begin{enumerate}
\item
Parse each such $N$  string into ${N}{L}$ vectors.

\item
Let $z_{1}^{L}(1)\in {\bf A}^L$ be the one L-vector that  satisfies the $d_{max}$ distortion criterion for the largest number of  L-vectors in the
incoming string and replace these L-vectors by $z_{1}^{L}(1)$.
\item
Let $z_{1}^{L}(2)\in {\bf A}^L$ be the one L-vector that  satisfies the $d_{max}$ distortion criterion for the largest number of the remaining, unreplaced   L-vectors in the $N$ string
 and replace these L-vectors by $z_{1}^{L}(2)$.
\item
In a similar way, generate $z_{1}^{L}(3)$, $z_{1}^{L}(4)$,... until all the L-vectors  in the $N$ sequence  are replaced.
\item
Sequentially feed the quantized  $N$ strings  into a version of the
constrained -dictionary LZ78(LRU) algorithm that  is described above, where now the alphabet consists of L-vectors in ${\bf A}^{L}$  rather than single letters in $A$, and where $D$ satisfies $\log D=L^{3}$ and
$N \geq D{\log D}$.

\end{enumerate}

The function  $p\log {\frac{1}{p}}$ is convex and
it's derivative,  $ \log {\frac{1}{p}} - \log{\frac{1}{e}}$ is positive for $0 \leq p \leq \frac{1}{e}$. Thus, for any $0 \leq p \leq \frac{1}{e}$:

$p\log {\frac{1}{p}} -(p-\delta)\log\frac{1}{p-\delta}\geq$
$\delta (\log {\frac{1}{p}} - p \log{\frac{1}{e}}$ for $0\leq \delta \leq p$.

Therefore, migrating L-vectors from any adaptive quantizer for L-vectors in the parsed input vector of length $N$, onto the  adaptive L-vectors quantizer that is described above yields, by it's majorization construction, an  empirical entropy
that is no larger than that of the  best adaptive L-vectors quantizer $\sum_{m=1}^{c} p(m){\log\left(\frac{1}{p(m)}\right)}$  plus a constant term $2\log{\frac{1}{e}}$.

Observe that the adaptive quantizer above is a finite-state machine with s(N) states where s(N)
 is  bounded by O($A^{2L}$) and where $s_o$=$s_i$ within the quantized sequence.

By Lemma 4, Theorem 1  and since $\frac{\log s}{L}$ vanishes as $D$ tends to infinity,

\begin{theorem}
The version LZ78(LRU) that is described above asymptotically achieves
$\FSLD(\bf x;d_{\max})$ as the dictionary size $D$ tends to infinity.
\end{theorem}

\end{document}